# Towards Lightweight and Privacy-preserving Data Provision in Digital Forensics for Driverless Taxi

Yanwei Gong, Xiaolin Chang, Jelena Mišić, Vojislav B. Mišić, Junchao Fan, Kaiwen Wang

*Abstract*—Data provision, referring to the data upload and data access, is one key phase in vehicular digital forensics. The unique features of *D*riverless *T*axi (DT) bring new issues to this phase: 1) efficient verification of data integrity when diverse *D*ata *P*roviders (DPs) upload data; 2) DP privacy preservation during data upload; and 3) privacy preservation of both data and *IN*vestigator (IN) under complex data ownership when accessing data.

To this end, we propose a novel *L*ightweight and *P*rivacy-preserving *D*ata *P*rovision (LPDP) approach consisting of three mechanisms: 1) the *P*rivacy-friendly *B*atch *V*erification *M*echanism (PBVm) based on elliptic curve cryptography, 2) *D*ata *A*ccess *C*ontrol *M*echanism (DACm) based on ciphertext-policy attribute-based encryption, and 3) *D*ecentralized *I*N *W*arrant *I*ssuance *M*echanism (DIWIm) based on secret sharing. Privacy preservation of data provision is achieved through: 1) ensuring the DP privacy preservation in terms of the location privacy and unlinkability of data upload requests by PBVm, 2) ensuring data privacy preservation by DACm and DIWIm, and 3) ensuring the identity privacy of IN in terms of the anonymity and unlinkability of data access requests without sacrificing the traceability. Lightweight of data provision is achieved through: 1) ensuring scalable verification of data integrity by PBVm, and 2) ensuring low-overhead warrant update with respect to DIWIm. Security analysis and performance evaluation are conducted to validate the security and performance features of LPDP.

## I. INTRODUCTION

As an important application of the Internet of things, intelligent transportation systems are evolving swiftly and then significantly enhancing the traveling convenience [1]. In recent years, there has been rapid advancement and deployment of driverless technology [2], exemplified by the introduction of driverless taxis (DTs) [3]. In contrast to manned vehicles, DT operations depend on an array of advanced sensors, control systems, and communication systems to gather environmental and status information, which is necessary for autonomously managing vehicle operation [4]. It can obtain the traffic information from the roadside units (RSU) [5] and service requests from the pedestrian apps [6]. Although DT utilizes artificial intelligence models and real-time data analysis to ensure driving safety [7], various safety accidents are inevitable [8]. When an accident occurs, conducting digital forensics for DT, known as vehicular digital forensics (VDF), becomes necessary to better determine liability in the accident [9].

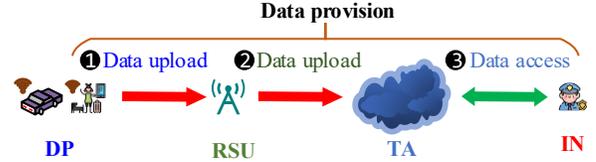

Fig.1. Data provision flow

VDF commonly consists of data upload and access (collectively referred to as **data provision** in this paper), data analysis, and result report phases [10]. As shown in Fig.1, the data provider (**DP**), such as in-vehicle data storage system, is responsible for collecting and uploading data to the trusted authority (**TA**), such as court. In order to avoid the time consumption caused by long-distance data transmission, DP can also upload the data not directly to the TA, but through RSU. The investigator (**IN**), such as police, can access and analyze the data after obtaining the warrant and then report the result [11]. Although there exist several VDF schemes, the data provision still faces the following three issues.

**I1. Data authenticity assurance under more diverse DPs.** Compared with manned vehicles, DTs cannot obtain more accident-related information from the driver after an accident. Therefore, it is crucial to obtain accident-related data from multiple DPs, such as RSU, pedestrian apps, in-vehicle data storage systems, etc. [12]. Due to the increasing types of DPs involved in data upload, how to ensure the data authenticity under more diverse DPs is an issue.

**I2. Efficient data upload with privacy protection under interactive real-time data generation.** The process of DT interacting with RSU and pedestrian apps generates a large amount of data, which is not all stored by DT [13]. Once an accident occurs, in order to ensure the integrity of data, it is necessary to upload these real-time data quickly. In addition, once pedestrian apps participate in the data upload, there is a risk that their data upload requests will expose their location privacy. Therefore, how to quickly upload the data with privacy protection under interactive real-time data generation is an issue.

**I3. Data privacy-preserving under complex data ownership.** The data uploaded by DPs has complex ownership. The data from the in-vehicle data storage system, pedestrian apps, and RSU is owned by the autonomous driving service provider, the DT service provider, and the traffic information provider, respectively [14]. Therefore, when accessing the data, how to protect data privacy by ensuring that IN can access the

Yanwei Gong, Xiaolin Chang, Junchao Fan and Kaiwen Wang are with the Beijing Key Laboratory of Security and Privacy in Intelligent Transportation, Beijing Jiaotong University, P.R.China. E-mail: {22110136, xlchang, 23111144, 24110146} @ bjtu.edu.cn.
  J. Mišić and V. B. Mišić are with Toronto Metropolitan University, Toronto ON, Canada M5B 2K3. E-mail: {jmisic, vmisic} @torontomu.ca.

data only if it obtains the access permission from the data owner is an issue.

With the above issues in mind, **to the best of our knowledge**, we propose the first *L*ightweight and *P*rivacy-preserving *D*ata *P*rovision (LPDP) approach in digital forensics for DT. It consists of three mechanisms, namely, 1) a *P*rivacy-friendly *B*atch *V*erification *M*echanism (PBVm) based on elliptic curve cryptography (ECC), 2) a *D*ata *A*ccess *C*ontrol *M*echanism (DACm) based on ciphertext-policy attribute-based encryption (CP-ABE), and 3) a decentralized IN warrant issuance mechanism (DIWIm) based on secret sharing. PBVm addresses the issuems of **I1** and **I2, and** DACm and DIWIm jointly tackle the issue of **I3**. The four distinctive security and performance features of LPDP are outlined below:

i. **Data authenticity and DP privacy-preserving when uploading data**. PBVm is applied for the DP authentication and data integrity verification so as to ensure data authenticity. PBVm can preserve DP privacy in terms of the DP location privacy and data upload request's unlinkability by allowing DP not to disclose its real identity.
ii. **Data and IN privacy-preserving with traceability when accessing data.** Both DACm and DIWIm are applied to ensures the data privacy-preserving. IN identity privacy is preserved by using the warrant in terms of the anonymity of IN and unlinkability of data access requests without sacrificing the traceability of malicious INs.
iii. **Scalable verification of data integrity**. LPDP applies PBVm to verify the integrity of a batch of data so as to reduce the computation overhead and then speed up data upload. That is, the increase in the scale of data for integrity verification brings a small increase in overhead, suggesting the scalability feature.
iv. **IN warrant update with low overhead**. LPDP ensures IN warrant update with low overhead by using DIWIm. Therefore, IN can only update the part of warrant instead of the whole warrant. To be specific, the complexity of warrant update reduces from $O(N)$ to $O(n)$, where $n$ is smaller than $N$ and both of them are the number of IN data access permission identifier.

We perform both formal and informal security assessments, to ensure that LPDP meets its declared security goals, including privacy-preserving authentication of DP and privacy-preserving of both IN and data. Performance evaluation is conducted to validate the lightweight feature of LPDF, which means that LPDF performs better than those in [24]-[26], and [28] in terms of computation and communication overheads.

The following describes the organization of the paper. Section II introduces the preliminaries. Section III reviews related work and describes the system. Section IV presents a detailed explanation of the proposed approach. Sections V and VI analyze the security and performance of LPDP, respectively. The paper concludes with Section VII.

## II. PRELIMINARIES

This section provides the cryptography and computationally hard problem used in LPDP. Specifically, elliptic curve, CP-ABE, and secret sharing are used to design PBVm, DACm, and DIWIm, respectively.

### A. Elliptic Curve

The equation $y^2 = x^3 + ax + b \mod p$ [15] defines an elliptic curve $E$ over a prime finite field $F_p$, where $a, b \in F_p$ and $4a^3 + 27b^2 \neq 0$. The set of all points on this curve, together with the point at infinity $O$, constitutes the elliptic curve group $\mathbb{G} = \{(x, y) : x, y \in F_p\}$, which supports operations such as scalar multiplication and point addition.

### B. Ciphertext-Policy Attribute-Based Encryption

CP-ABE encryption scheme consists of four algorithms: Setup, Encrypt, KeyGen, and Decrypt [16]:

- $Setup(\kappa, \mathbb{A})$: This algorithm generates a public key *PK* and the corresponding secret key *SK* with the input of the security parameter $\kappa$ and the attributes set $\mathbb{A}$.
- $Encrypt(\mathbb{P}, MPK, m)$: This algorithm generates a ciphertext *C* with the input of the access policy $\mathbb{P}$, the public key *PK* and the plaintext *m*.
- $KeyGen(\mathbb{A}, MPK, MSK)$: This algorithm generates a user secret key $k_u$ about $\mathbb{A}$ with the input of the attribute set $\mathbb{A}$, the public key *PK* and the secret key *SK*.
- $Decrypt(\mathbb{P}, MPK, k_u, C)$: This algorithm generates the plaintext *m* or null with the input of the ciphertext *C*, the access policy $\mathbb{P}$, the public key *PK* and the secret key $k_u$ about the attribute set $\mathbb{A}$.

### C. Secret Sharing

A secret sharing scheme [17] over $Z_q$ consists of a tuple of algorithms $(A_1, A_2)$.

- $A_1(n, t, s)$: This algorithm creates shares $(s_1, ..., s_n)$ of the secret $s$. The threshold $t$ is set such that $0 < t \leq n$.
- $A_2(s_1, ..., s_t)$: This algorithm recovers $s$ using the Lagrange's interpolation formula.

### D. Computationally Hard Problem

We list three computationally difficult problems on which LPDP's security is based.

1) Elliptic Curve Discrete Logarithm Problem (ECDLP) [18]: Given $X, Y \in \mathbb{G}$, it is computationally hard for any polynomial-time bounded algorithm to find $y$ such that $y \in Z_p^*$ and $Y = yX$.
2) Elliptic Curve Diffie-Hellman Problem (ECDHP) [19]: Given $P \in \mathbb{G}, x, y \in Z_p^*$, it is computationally hard for any polynomial-time bounded algorithm to find $xyP$ such that $X = xP, Y = yP$.
3) *q*-Generalized Diffie-Hellman Problem (q-GDHP) [20]: Given $a_1 P, a_2 P ..., a_q P$ in $\mathbb{G}$ and all the subset products

$(\prod_{i \in S} a_i)P \in \mathbb{G}$ for any strict subset $S \subset \{1,...,q\}$, it is computationally hard for any polynomial-time bounded algorithm to compute $(a_1...a_q)P \in \mathbb{G}$.

## III. RELATED WORK AND SYSTEM DESCRIPTION

In this section, related work and system description are provided. TABLE I presents the comparison of related work.

### A. Related Work

There are already scholars devoting themselves to VDF. Li et al. [21] proposed a blockchain-based digital forensics scheme, supporting DP privacy protection by utilizing short randomizable signatures. CP-ABE based on bilinear pair (BP) was also used to realize fine-grained access control. Due the large overhead caused by using BP, they further designed another blockchain-based VDF scheme [25]. This scheme protected the privacy of IN by using decentralized anonymous credentials and zero-knowledge proof, which also brought a large overhead. Ranu et al. [22] proposed an intelligent digital forensics system based on blockchain for vehicles. They utilized randomizable signatures and CP-ABE to realize the privacy protection for DP and flexible access control, respectively. A prototype was implemented to evaluate the performance of the proposed system. Li et al. [23] designed a VDF scheme focusing on the traceability while maintaining flexible access to data based on key-policy attribute-based encryption (KP-ABE). However, the KP-ABE used in this scheme was also based on BP, which brought a large overhead. Zhang et al. [24] designed an authentication scheme for VDF services. A three-tier architecture was proposed to realize the secure anonymous authentication by using pairing-free certificateless signcryption. However, the scheme only considered the authentication and privacy protection for DP. Li et al. [26] proposed an anonymous and secure digital-forensics scheme for vehicles. A trusted execution environment was used to boost secure data upload and request. Chen et al. [27] designed a digital-forensic model for vehicular terminal devices and presented forensic cases in order to discuss the model's applicability and validity. Li et al. [28] proposed a VDF scheme focusing on data integrity verification by using proxy re-encryption (PRE). However, they only considered in-vehicle digital forensics.

These existing schemes are only capable of digital forensics for manned vehicles. They didn't consider three new issues, which are introduced in Section I. But we propose LPDP for DT by addressing these issues.

TABLE I COMPARISON OF RELATED WORK

| Ref. | DP authentication | DP privacy protection | Data privacy protection | Decentralized warrant issuance | IN traceability | Data integrity batch verification | Data access control Algorithm | Data access control Cryptographic primitive | Lightweight |
|---|---|---|---|---|---|---|---|---|---|
| [21] Li et al. 2021 | √ | √ | √ | × | × | × | CP-ABE | BP | × |
| [22] Ranu et al. 2022 | √ | √ | √ | × | × | × | CP-ABE | BP | × |
| [23] Li et al. 2022 | √ | × | √ | √ | √ | × | KP-ABE | BP | × |
| [24] Zhang et al. 2023 | √ | × | √ | × | × | × | × | × | √ |
| [25] Li et al. 2023 | √ | × | √ | √ | √ | × | CP-ABE | BP | × |
| [26] Li et al. 2024 | √ | × | √ | × | √ | × | CP-ABE | BP | × |
| [27] Chen et al. 2024 | √ | × | √ | × | × | × | × | × | × |
| [28] Li et al. 2024 | √ | × | √ | × | × | × | PRE | ElGamal | × |
| Ours 2024 | √ | √ | √ | √ | √ | √ | CP-ABE | ECC | √ |

### B. System Description

This section provides an overview of the system, covering its entities, design goals, and security model. System entities describe which entities are included in LPDP and their responsibilities and characteristics. The design goals list the goals that need to be achieved in order to ensure that LPDP is lightweight and privacy-preserving. The security model, which is for LPDP's two important security attributes: unforgeability and confidentiality, is defined by using cryptographic games. Both attributes are important for LPDP to ensure design goals.

### B.1 System Entity

This section describes system entities, including the trusted authority (TA), data provider (DP), warrant issuer (WI), roadside unit (RSU), and investigator (IN). The system framework is illustrated in Fig.2.

**Trusted Authority (TA)**: TA is responsible for generating the public parameters used in the registration of DP, WI, RSU, and IN by generating public-private key pairs for them. Besides, TA also has its own public-private key pair. Moreover, TA will respond the data access requests sent by IN and return the requested data, which is stored in a distributed data storage system.

**Data provider (DP)**: DP takes responsibility for uploading data related to accident. In reality, DP can be DT in-vehicle data storage system, pedestrian apps, or other DTs, all of which are resource-constrained. Before being able to upload data, DP must register via TA and obtain its identity and data access policy used in the data upload sub-phase.

**Warrant Issuer (WI)**: WI is in charge of issuing the warrant for INs. It consists of several institutions, such as the autonomous driving service provider, the DT service provider, the traffic information provider. According to the data access permission identifier of IN, each WI issues a part of warrant for it. The access permission identifier can be either 1 or 0, representing whether access is allowed or denied, respectively. Before being able to issue the warrant, each WI must register via TA and obtain its public-private key pair.

**Roadside Unit (RSU)**: RSU is used to assist DP to upload data to the distributed data storage system, which stores the data related to accidents. When an accident occurs, RSU communicates with DP and accepts the uploaded data after authenticating DP. Then, RSU further uploads the data to the distributed data storage system. RSU must register via TA and obtain its public-private key pair, too.

**Investigator (IN)**: IN is responsible for data access and analysis of accidents. Before that, it must complete the registration via TA and obtain the warrant from WIs. When an accident occurs, it can send the data access request to TA and obtain the accident-related data.

*B.2 Design Goals*

To ensure the privacy protection and lightweight features of LPDP, we outline the design goals with respect to security and performance. These goals will be used for the informal security and performance analyses presented in Sections V.B and VI.C, respectively.

*B.2.1 Security Goals*

The descriptions of security goals are as follows:

**S1) Data authenticity**. LPDP must assure the authenticity of the data uploaded by DP and the data accessed by IN from the aspects of data origin's authenticity and data integrity.

**S2) Data privacy protection**. During the data access process, LPDP implements the minimal disclosure of data. Specifically, different INs are only able to access the data assigned to them instead of all the data.

**S3) Privacy protection of DP location**. When a DP uploads the data, attackers can't obtain the location information of this DP.

**S4) Decentralized warrant issuance**. In LPDP, the issuance of IN warrant is not done by a trusted third party.

**S5) Anonymity**. When accessing data, attackers can't obtain the real identity of IN.

**S6) Unlinkability**. The data upload requests sent by DP are unlinkable to attackers. The data access requests sent by IN are also unlinkable to attackers.

**S7) Traceability**. When IN has misbehaviors, such as leaking data, TA can trace this malicious IN.

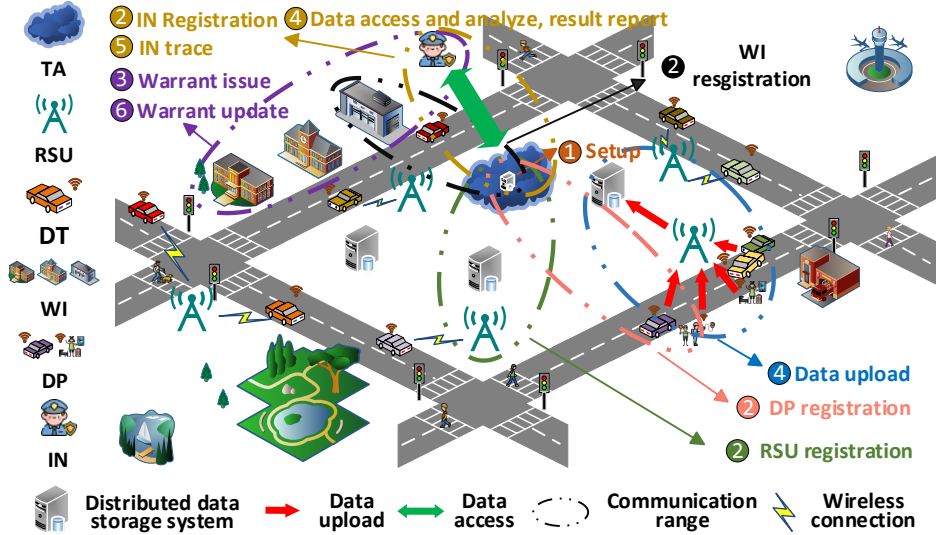

Fig.2. System framework

*B.2.2 Performance Goals*

There are three performance goals are as follows:

**P1) Warrant update with low overhead**. When IN needs to update the warrant, the overhead of warrant update should be less than the overhead of reapplying for a warrant.

**P2) Efficient verification of data integrity**. When verifying data integrity, the extra overhead caused by the increase in the scale of data for integrity verification increases slowly.

*B.3 Security Model*

In addition to evaluating the privacy protection of LPDP by establishing security goals, we define the security model concerning LPDP's unforgeability and confidentiality through cryptographic games. These games facilitate the formal security analysis of LPDP using the formal security analysis tool [32], which is detailed in Section V.A. To be specific, *Game Unforgeability* (GU) and *Game Confidentiality* (GC) are presented to simulate the interaction of entities in LPDP, both of which contains many queries sent by any probabilistic polynomial time (PPT) **adversary** $\mathcal{A}$ to the **challenger** $\mathcal{C}$. For clarity, the symbols used in this section are listed in TABLE II.

*B.3.1 Game Unforgeability*

We first give the fundamental descriptions of GU and then define the unforgeability in **Definition 1**:

1) Setup: $\mathcal{C}$ generates the public parameters $pp = \{E, \mathbb{G}, G, p, pk_{war}, pk_{TA}, H_1, H_2, H_3\}$ under the security parameters $\kappa$.

2) Adversary queries: The queries are sent by $\mathcal{A}_1$ to $\mathcal{C}$, which are described as follows:
   a) Public-private key pair query: $\mathcal{A}_1$ sends public-private key pair query to $\mathcal{C}$. $\mathcal{C}$ randomly selects $sk_{\mathcal{A}_1} \leftarrow Z_p$, computes $pk_{\mathcal{A}_1} = sk_{\mathcal{A}_1} G$ and returns

$(pk_{\mathcal{A}_I}, sk_{\mathcal{A}_I})$.

b) Token query: $\mathcal{A}_I$ sends token query to $\mathcal{C}$. $\mathcal{C}$ randomly selects $t \leftarrow Z_p$ returns it.

c) Signature query: $\mathcal{A}_I$ sends $m$ to $\mathcal{C}$. $\mathcal{C}$ generates the signature $sig_m$ returns it.

For the hash functions $H_1$ used in this game, $\mathcal{C}$ defines corresponding query for it as follows:

- $H_1$ oracle query: If input $t$, $\mathcal{C}$ randomly chooses $t_h \leftarrow Z_p$ and returns.

When the game ends, $\mathcal{A}_I$ outputs a signature $sig_{m'}$ for $m'$ by using $(pk_{\mathcal{A}_I}, sk_{\mathcal{A}_I})$ and $t$, all of which are never be queried. Then, if $sig_m$ is a valid signature of $m'$, $\mathcal{A}_I$ wins the game.

**Definition 1**. **Unforgeability**. For $\mathcal{A}_I$, if $Adv_{\mathcal{A}_I}^{LPDP} = |\Pr(win_{\mathcal{A}_I}^{GU})| < \varepsilon$ holds, where $GU$ is *Game Unforgeability* and $\varepsilon$ is negligible, then LPDP satisfies the unforgeability.

*B.3.2 Game Confidentiality*

We first give the fundamental descriptions of GC and then define the confidentiality in **Definition 2**:

1) Setup: $\mathcal{C}$ generates the public parameters $pp = \{E, \mathbb{G}, G, p, pk_{war}, pk_{TA}, H_1, H_2, H_3\}$ under the security parameters $\kappa$.

2) Adversary queries: $\mathcal{A}_{II}$ sends queries to $\mathcal{C}$, which are described as follows:

   a) Public-private key pair query: $\mathcal{A}_{II}$ sends public-private key pair query to $\mathcal{C}$. $\mathcal{C}$ randomly selects $sk_{\mathcal{A}_{II}} \leftarrow Z_p$, computes $pk_{\mathcal{A}_{II}} = sk_{\mathcal{A}_{II}} G$ and returns $(pk_{\mathcal{A}_{II}}, sk_{\mathcal{A}_{II}})$.

   b) Warrant issue query: $\mathcal{A}_{II}$ sends warrant issue query to $\mathcal{C}$. $\mathcal{C}$ generates $w_{\mathcal{A}_{II}} = \{w_{1,i}, w_{2,i}, w_{3,i}, w_{4,i}\}_{i=1}^{N_{WI}}$ and returns.

   c) Decryption query: $\mathcal{A}_{II}$ sends $\{\mathbb{P}, \{L_i\}_{i=1}^{N_{WI} - |\mathbb{P}|}, N_1, N_2, c_1, c\}$, which is the ciphertext of $m$, to $\mathcal{C}$. $\mathcal{C}$ returns $m$.

For the hash functions $H_1, H_2, H_3$ used in this game, $\mathcal{C}$ defines corresponding queries for them as follows:

1) $H_1$ oracle query: If input $(\mathbb{P}, m', k')$, $\mathcal{C}$ randomly chooses $v_1 \leftarrow Z_p$ and returns.

2) $H_2$ oracle query: If input $KDF(n_1 G)$ or $k^{A'}$, $\mathcal{C}$ randomly chooses $v_2 \leftarrow \{0,1\}^*$ and returns.

3) $H_3$ oracle query: If input $i$, $\mathcal{C}$ randomly chooses $v_3 \leftarrow Zp$ and returns.

When the game ends, $\mathcal{A}_{II}$ sends $m_0, m_1$ to $\mathcal{C}$. Then, $\mathcal{C}$ randomly picks $b \leftarrow \{0,1\}$ and returns $\{\mathbb{P}, \{L_i\}_{i=1}^{N_{WI} - |\mathbb{P}|}, N_1, N_2, c_1, c_{m_b}\}$ to $\mathcal{A}_{II}$. Then $\mathcal{A}_{II}$ outputs the $b'$. If $b' = b$, $\mathcal{A}_{II}$ wins the game.

**Definition 2**. **Confidentiality**. For $\mathcal{A}_{II}$, if $Adv_{\mathcal{A}_{II}}^{LPDP} = |\Pr(win_{\mathcal{A}_{II}}^{GC}) - \frac{1}{2}| < \varepsilon$ holds, where $GC$ is *Game Confidentiality* and $\varepsilon$ is negligible, then LPDP satisfies the confidentiality.

IV. THE CONSTRUCTION OF LPDP

In this section, the details of LPDP, consisting of setup registration, warrant issue, data provision, traceability, and warrant update phases, are introduced. The symbols used are presented in TABLE II.

TABLE II  SYMBOLS AND DESCRIPTION

| Symbol | Description |
| --- | --- |
| KDF | Key derivation function |
| $\kappa$ | Security parameters of LPDP |
| $pp$ | Public parameters of LPDP |
| $|G|$ | The size of generator $G$ |
| $|p|$ | The size of prime order $p$ |
| $(pk_{XX}, sk_{XX})$ | Public-private key pair of XX |
| $ID_{XX}$ | Real identity of XX |
| $PID_{XX}$ | Pseudonym of XX |
| $H_1, H_2, H_3$ | Hash function |
| $t$ | Identity token of DP for uploading data |
| $X^A$ | The data related to the vehicle digital forencisc of accident $A$ |
| $B_{DP}^A$ | The number of signature in a batch verification for accident $A$ |
| $\mathbb{D}$ | The data access permission identifier set |
| $\mathbb{A}$ | The data access permission identifier set of IN |
| $\mathbb{P}$ | The data access policy |
| $|\mathbb{A}|$ | The number of 1 in $\mathbb{A}$ |
| $|\mathbb{P}|$ | The number of 1 in $\mathbb{P}$ |
| $|X|$ | The number of 1 in string $X$ |
| $w_{IN}$ | The warrant of IN |
| $N_{WI}$ | The number of WIs |
| $N_U$ | The number of updated partial warrant |
| $req_{XX}$ | The request of XX |
| $Enc(m)_{pk}$ | Encrypt m by *pk* using ECC |
| $Sig(m)_{sk}$ | Sign m by *sk* using ECC |
| $\Pr(win_{\mathcal{A}}^{Game})$ | The probability of $\mathcal{A}$ wins the *Game* |

*A. Setup Phase*

In this phase, given $\mathbb{P}$, $\kappa$, and $\mathbb{D} = \{a_i\}_{i=1}^{N_{WI}} (a_i \in \{0,1\})$, TA will generate the public parameters by the following five steps:

1) TA selects an elliptic curve $E: y^2 = x^3 + ax + b \bmod p$ and obtains the group $\mathbb{G}$ of prime order $p$ with a generator $G$.

2) TA first randomly selects $\{s_1, \{s_{2,i}\}_{i=1}^{N_{WI}}, s_3\} \leftarrow Z_p$, computes $s_2 = s_{2,1} \cdot \ldots \cdot s_{2,N_{WI}}$, $X_i = s_1^i G$, $Y_i = s_1^i s_2 G$ and $Z_i = s_1^i s_3 G$ for $i \in \{1,\ldots,N_{WI}\}$, and then obtains both $sk_{war} = \{s_1, \{s_{2,i}\}_{i=1}^{N_{WI}}\}, s_2, s_3\}$ and $pk_{war} = \{X_i, Y_i, Z_i\}_{i=1}^{N_{WI}}$.

3) TA randomly selects $t, sk_{TA} \leftarrow Z_p$ and then computes $pk_{TA} = sk_{TA} G$.

4) TA chooses three hash functions: $H_1: \{0,1\}^* \to Z_p$, $H_2: \{0,1\}^* \to \{0,1\}^*$, and $H_3: \{0,1\}^* \to Z_p$.

5) TA publishes the public parameters $pp = \{E, \mathbb{G}, G, p, pk_{war}, pk_{TA}, H_1, H_2, H_3\}$.

### B. Registration Phase

In this phase, DPs, WIs, RSUs and INs will complete their registration by sending their real identities to TA.

#### B.1 WI Registration

$\{WI_i\}_{i=1}^{N_{WI}}$ will complete registration via TA. $WI_i$ first computes $Enc(ID_{WI_i})_{pk_{TA}}$ and sends the result to TA.

After receiving the message, TA uses the following three steps to generate the public-private key pair for $WI_i$.

1) TA randomly selects $sk_{WI_i} \leftarrow Z_p$ and computes $pk_{WI_i} = sk_{WI_i} G$.

2) TA chooses a polynomial $f(x) = s_3 + b_1 x + ... + b_{N-1} x^{N-1}$ with degree $N_{WI} - 1$ and computes $s_{3,i} = H_1(ID_{WI_i})$, $f(s_{3,i})$.

3) TA sends $\{pk_{WI_i}, sk_{WI_i}, s_1, s_2, s_{2,i}, s_{3,i}, f(s_{3,i})\}$ to $WI_i$ via a secure channel and publishes $\{s_{3,i}\}_{i=1}^{N_{WI}}$.

So far, $WI_i$ completes its registration.

#### B.2 DP Registration

DP will also complete registration via TA by computing $Enc(ID_{DP})_{pk_{TA}}$ and then sending the result to TA.

After receiving the message, TA generates the public-private key pair for DP through the following two steps.

1) TA randomly selects $sk_{DP}, t \leftarrow Z_p$ and computes $pk_{DP} = sk_{DP} G$.

2) TA sends $\{pk_{DP}, sk_{DP}, t, \mathbb{P}\}$ to DP via a secure channel, where $\mathbb{P}$ is the access policy used to encrypt and upload data.

So far, DP completes its registration.

#### B.3 RSU Registration

RSU will complete registration via TA by computing $Enc(ID_{RSU})_{pk_{TA}}$ and then sending the result to TA.

After receiving the message, TA generates the public-private key pair for RSU through the following two steps.

1) TA randomly selects $r_{RSU} \leftarrow Z_p$ and computes $pk_{RSU} = r_{RSU} G$, $sk_{RSU} = sk_{TA} H_1(t) + r_{RSU}$.

2) TA sends $\{pk_{RSU}, sk_{RSU}\}$ to RSU via a secure channel.

So far, RSU completes its registration.

#### B.4 IN Registration

IN will complete registration via TA by computing $Enc(ID_{IN})_{pk_{TA}}$ and then sending the result to TA.

After receiving the message, TA generates the public-private key pair for IN through the following two steps:

1) TA randomly selects $sk_{IN} \leftarrow Z_p$ and computes $pk_{IN} = sk_{IN} G$, $PID_{IN} = H_1(ID_{IN})$.

2) TA sends $\{pk_{IN}, sk_{IN}, PID_{IN}\}$ to IN via a secure channel.

So far, IN completes its registration.

### C. Warrant Issue Phase

In the warrant issue phase, IN will apply WIs for the warrant, which is used in the data provision phase. The process of issuing warrant uses DIWIm. To be specific, when applying the warrant, each WI will issue a partial warrant for IN. Then IN can obtain the complete warrant by aggregating these partial warrants. Besides, due to the different data access permission identifiers of different INs, they will also obtain the different warrants, determining the data they can access. Assuming that the data access permission identifier set of IN is $\mathbb{A} = \{a_{WI_i}\}_{i=1}^{N_{WI}}$ ($a_{WI_i} \in \{0,1\}$), the two steps are described as follows:

1) IN computes $Sig(req_{warrant}, T)_{sk_{IN}}$ where $req_{warrant}$ and $T$ represent the request for the warrant and timestamp, respectively.

2) IN sends $\{Sig(req_{warrant})_{sk_{IN}}, req_{warrant}, T\}$ to $\{WI_i\}_{i=1}^{N_{WI}}$.

After receiving the message, $WI_i$ verifies $Sig(req_{warrant})_{sk_{IN}}$. If true, it conducts the following four steps:

1) Each $WI_i$ computes $y_i = s_1 + H_3(i)^{1-a_{WI_i}}$, where $a_{WI_i}$ is the data access permission identifier of IN stored in $WI_i$.

2) Each $WI_i$ randomly selects $d_{1,i}, d_{2,i} \leftarrow Z_p$ and sends $Enc(d_{1,j}, d_{2,j})_{sk_{WI_j}}$ to other $WI_j$ for $j \in [1, N_{WI}], j \neq i$.

3) Each $WI_i$ computes $d_1 = \prod_{i=1}^{N_{WI}} d_{1,i}$, $d_2 = \prod_{i=1}^{N_{WI}} d_{2,i}$, $w_{2,i} = -d_2 f(s_{3,i}) \prod_{j=1, j\neq i}^{N_{WI}} (-s_{3,j}/(s_{3,i} - s_{3,j}))$, $w_{1,i} = d_1 + s_2 d_2$ and $w_{4,i} = -d_1 f(s_{3,i}) (\prod_{j=1, j\neq i}^{N_{WI}} -s_{3,j}/(s_{3,i} - s_{3,j}))$ $/s_2$. Then, $WI_i$ obtains $w_{IN_i} = (w_{1,i}, w_{2,i}, w_{3,i}, w_{4,i})$.

4) Each $WI_i$ sends $\{enc(w_{IN_i}, a_{WI_i})_{pk_{IN}}, sig(T)_{sk_{WI_i}}\}$ to IN.

After receiving the message, IN verifies $sig(T)_{sk_{WI_i}}$. If true, IN conducts the following two steps to obtain the complete warrant:

1) IN verifies $w_{1,1} = ... = w_{1,N_{WI}}$. If true, it continues to compute $w_1 = w_{1,1} = ... = w_{1,N_{WI}}$, $w_2 = \sum_{i=1}^{N_{WI}} w_{2,i}$, $w_3 = \prod_{i=1}^{N_{WI}} w_{3,i}$, and $w_4 = \sum_{i=1}^{N_{WI}} w_{4,i}$.

2) IN computes $w_{IN,1} = w_1$, $w_{IN,2} = w_2 + w_3 + w_4$, and obtains the complete warrant $w_{IN} = (w_{IN,1}, w_{IN,2})$ and $\mathbb{A} = \{a_{WI_i}\}_{i=1}^{N_{WI}}$.

### D. Data Provision Phase

In the data provision phase, IN will access data when an accident occurs. Before that, DPs will upload the data related to the accident via RSU. Assume that the occurred vehicle accident is $A$.

*D.1 Data Upload*

To ensure the data authenticity, DP will be authenticated by RSU and data integrity will also be verified by RSU. In order to enable fast DP authentication and data integrity verification to speed up data upload, we design PBVm. Assuming that there are $N_{DP}^A$ DPs for $A$ and the data access policy is $\mathbb{P}=\{p_i\}_{i=1}^{N_{WI}}$. Taking $DP_i$ for example, the whole four-step process of data upload is as follows:

1) $DP_i$ randomly selects a number $k_i^A \leftarrow \{0,1\}^*$ and computes $n_{1,i}^A = H_1(\mathbb{P}, M, k_i^A)$, $n_{2,i}^A = KDF(n_{1,i}^A G)$.

2) $DP_i$ computes $\{L_{i,j}^A = n_{1,i}^A X_j\}_{j=1}^{N_{WI}-|\mathbb{P}|}$, $N_{1,i}^A = n_{1,i}^A \sum_{j=1}^{N_{WI}} g_{1,j} Y_j$, $N_{2,i}^A = n_{1,i}^A \sum_{j=1}^{N_{WI}} g_{1,j} Z_j$, $c_{1,i}^A = H_2(n_{2,i}^A) \oplus k_i^A$, and $c_{2,i}^A = H_2(k_i^A) \oplus m_i^A$, where $x^j$'s coefficient in $g_1(x,\mathbb{P}) = \prod_{j=1}^{N_{WI}}(x+H_3(j))^{1-p_j}$ is $g_{1,j}$.

3) $DP_i$ randomly selects a number $r_{token,i}^A \leftarrow Z_p$ and computes $R_i^A = r_{token,i}^A G$. Besides, it computes $sig_i^A = r_{token,i}^A(pk_{RSU} + H_1(t)pk_{TA})H_1(c_{2,i}^A, A)$.

4) $DP_i$ sends $\{\{L_{i,j}^A\}_{j=1}^{N_{WI}-|\mathbb{P}|}, N_{1,i}^A, N_{2,i}^A, c_{1,i}^A, c_{2,i}^A, sig_i^A, R_i^A\}$ to RSU.

After receiving the message from multiple $\{DP_i\}_{i=1}^{B_{DP}^A}(1 \le B_{DP}^A \le N_{DP}^A)$, RSU conducts the following two steps before uploading data:

1) RSU computes $sig^A = \sum_{i=1}^{B_{DP}^A} sig_i^A$.

2) RSU verifies if $sig^A sk_{RSU}^{-1} = \sum_{i=1}^{B_{DP}^A} R_i^A H_1(c_{2,i}^A, A)$ holds. If true, RSU accepts $\{\{L_{i,j}^A\}_{j=1}^{N_{WI}-|\mathbb{P}|}, N_{1,i}^A, N_{2,i}^A, c_{1,i}^A, c_{2,i}^A\}_{i=1}^{B_{DP}^A}$ and uploads them to the distributed data storage system.

So far, the data have already been uploaded.

**Remark 1. Correctness of batch verification.** We provide the correctness analysis of batch verification as follows:

$$sig^A sk_{RSU}^{-1} = sk_{RSU}^{-1} \sum_{i=1}^{B_{DP}^A} sig_i^A$$
$$= sk_{RSU}^{-1} \sum_{i=1}^{B_{DP}^A} r_{token,i}^A (pk_{RSU} + H_1(t)pk_{TA}) H_1(c_{2,i}^A, A)$$
$$= sk_{RSU}^{-1} \sum_{i=1}^{B_{DP}^A} r_{token,i}^A (r_{RSU} \cdot G + (sk_{RSU} - r_{RSU})G) H_1(c_{2,i}^A, A)$$
$$= sk_{RSU}^{-1} \sum_{i=1}^{B_{DP}^A} r_{token,i}^A sk_{RSU} G H_1(c_{2,i}^A, A)$$
$$= \sum_{i=1}^{B_{DP}^A} r_{token,i}^A G H_1(c_{2,i}^A, A) = \sum_{i=1}^{B_{DP}^A} R_i^A H_1(c_{2,i}^A, A)$$

*D.2 Data Access*

After the data uploaded, IN will send the data access request to TA for data analysis. In order to protect data privacy, DACm is used in the process of data access. First, IN computes $\{Enc(req_{data}, PID_{IN}, T)_{pk_{TA}}\}$ and sends the result to TA.

When receiving message, TA conducts the following two steps:

1) TA decrypts $Enc(req_{data}, PID_{IN}, T)_{pk_{TA}}$ and obtains $PID_{IN}$.

2) TA computes $c^A = c_2^A \oplus (PID_{IN}, sk_{IN})$ and sends data $\{\mathbb{P}, \{L_i^A\}_{i=1}^{N_{WI}-|\mathbb{P}|}, N_1^A, N_2^A, c_1^A, c^A\}$ to IN.

When receiving the data from TA, IN decrypts the data through the following four steps:

1) IN computes $\{h_i = a_i - p_i\}_{i=1}^{N_{WI}}$, $Y^{A'} = w_{IN,2} N_1^A$, and $Z^{A'} = w_{IN,1} N_2^A$.

2) IN computes $K^A = \sum_{i=1}^{N_{WI}-|\mathbb{P}|} g_{2,i} L_i^A$ and $n_1^A G = (Y^{A'} + Z^{A'} - K^A) / \prod_{i=1}^{N_{WI}-|\mathbb{P}|} H_3(i)^{h_i}$, where $x^i$'s coefficient in $g_2(x, \mathbb{P}, \mathbb{A}) = g_1(x, \mathbb{P}) / g_1(x, \mathbb{A})$ is $g_{2,i}$.

3) IN computes $k^{A'} = H_2(KDF(n_1^A G)) \oplus c_1^A$, $m^{A'} = H_2(k^{A'}) \oplus c^A \oplus (PID_{IN} sk_{IN})$, and $n_1^{A'} = H_1(\mathbb{P}, m^{A'}, k^{A'})$.

4) IN verifies if $n_1^{A'} G = n_1^A G$ holds. If true, it accepts $m^{A'}$.

So far, IN obtains the data and can analyze it.

**Remark 2. Correctness of data access.** We provide the correctness analysis of data access as follows:

$$m^{A'} = H_2(k^{A'}) \oplus c^A \oplus (PID_{IN} sk_{IN})$$
$$= H_2(k^{A'}) \oplus c_2^A \oplus (PID_{IN} sk_{IN}) \oplus (PID_{IN} sk_{IN})$$
$$= H_2(H_2(KDF((Y^{A'} + Z^{A'} - K^A) / \prod_{i=1}^{N_{WI}-|\mathbb{P}|} H_3(i)^{h_i})) \oplus c_1^A)$$
$$\oplus c_2^A$$
$$= H_2(H_2(KDF(w_{IN,2} N_1^A + w_{IN,1} N_2^A - K^A) /$$
$$\prod_{i=1}^{N_{WI}-|\mathbb{P}|} H_3(i)^{h_i})) \oplus c_1^A) \oplus c_2^A$$

The $w_{IN,1}$ and $w_{IN,2}$ are obtained in the warrant issue phase. Therefore, we have

$$w_{IN,2} N_1^A + w_{IN,1} N_2^A$$
$$= (w_2 + w_3 + w_4) N_1^A + w_1 N_2^A$$
$$= (\sum_{i=1}^{N_{WI}} w_{2,i} + \prod_{i=1}^{N_{WI}} w_{3,i} + \sum_{i=1}^{N_{WI}} w_{4,i}) N_1^A + w_1 N_2^A$$
$$= (\sum_{i=1}^{N_{WI}} -d_2 f(s_{3,i}) \prod_{j=1,j\ne i}^{N_{WI}} \frac{-s_{3,j}}{s_{3,i}-s_{3,j}} + \prod_{i=1}^{N_{WI}} \frac{1}{s_{2,i} y_i}$$
$$+ \frac{\sum_{i=1}^{N_{WI}} -d_1 f(s_{3,i}) \prod_{j=1,j\ne i}^{N_{WI}} \frac{-s_{3,j}}{s_{3,i}-s_{3,j}}}{s_2}) N_1^A + w_1 N_2^A$$
$$= (-d_2 s_3 + \prod_{i=1}^{N_{WI}} \frac{1}{s_{2,i} y_i} - \frac{d_1 s_3}{s_2}) N_1^A + w_1 N_2^A$$
$$= (\prod_{i=1}^{N_{WI}} \frac{1}{s_{2,i} s_1 + H_3(i)^{1-a_{W_i}}} - d_2 s_3 - \frac{d_1 s_3}{s_2}) N_1^A + w_1 N_2^A$$
$$= (\frac{1}{s_2 \prod_{i=1}^{N_{WI}} s_1 + H_3(i)^{1-a_{W_i}}} - d_2 s_3 - \frac{d_1 s_3}{s_2}) N_1^A + w_1 N_2^A$$
$$= (\frac{1}{s_2 \prod_{i=1}^{N_{WI}} s_1 + H_3(i)^{1-a_{W_i}}} - d_2 s_3 - \frac{d_1 s_3}{s_2}) N_1^A + (d_1 + s_2 d_2) N_2^A.$$

Therefore, we have

$w_{IN,2}N_1^A + w_{IN,1}N_2^A - K^A$

$= (\dfrac{1}{s_2\prod_{i=1}^{N_{WI}}(s_1+H_3(i)^{1-a_{WI_i}})} - d_2s_3 - \dfrac{d_1s_3}{s_2})(n_1^A\sum_{j=1}^{N_{WI}}g_{1,j}Y_j) +$

$(d_1 + s_2d_2)(n_1^A\sum_{j=1}^{N_{WI}}g_{1,j}Z_j) - K^A$

$= (\dfrac{1}{s_2\prod_{i=1}^{N_{WI}}(s_1+H_3(i)^{1-a_{WI_i}})} - d_2s_3 - \dfrac{d_1s_3}{s_2})n_1^As_2g_1(s_1,\mathbb{P}) +$

$(d_1 + s_2d_2)n_1^As_3g_1(s_1,\mathbb{P}) - K^A$

$= ((\dfrac{1}{s_2\prod_{i=1}^{N_{WI}}(s_1+H_3(i)^{1-a_{WI_i}})} - \dfrac{d_1s_3}{s_2})n_1^As_2 - d_2s_3n_1^As_2 + d_1n_1^As_3$

$-s_2d_2n_1^As_3)g_1(s_1,\mathbb{P})G - K^A$

$= n_1^Ag_1(s_1,\mathbb{P})((\dfrac{1}{s_2\prod_{i=1}^{N_{WI}}(s_1+H_3(i)^{1-a_{WI_i}})} - \dfrac{d_1s_3}{s_2})s_2 + d_1s_3)G - K^A$

$= n_1^A(g_1(s_1,\mathbb{P})/g_1(s_1,\mathbb{A}))G - (\sum_{i=1}^{N_{WI}-|\mathbb{P}|}g_{2,i}s_1^i))G$

$= n_1^A(g_2(s_1,\mathbb{P},\mathbb{A}) - \sum_{i=1}^{N_{WI}-|\mathbb{P}|}g_{2,i}s_1^i + \prod_{i=1}^{N_{WI}-|\mathbb{P}|}H_3(i)^{h_i} -$

$\prod_{i=1}^{N_{WI}-|\mathbb{P}|}H_3(i)^{h_i})G$

$= n_1^A(g_2(s_1,\mathbb{P},\mathbb{A}) - g_2(s_1,\mathbb{P},\mathbb{A}) + \prod_{i=1}^{N_{WI}-|\mathbb{P}|}H_3(i)^{h_i})G$

$= n_1^A(\prod_{i=1}^{N_{WI}-|\mathbb{P}|}H_3(i)^{h_i})G$.

Therefore, we have

$m^{A'} = H_2(H_2(KDF((n_1^A\prod_{i=1}^{N_{WI}-|\mathbb{P}|}H_3(i)^{h_i}/\prod_{i=1}^{N_{WI}-|\mathbb{P}|}H_3(i)^{h_i})G))$

$\oplus c_1^A) \oplus c_2^A$

$= H_2(H_2(KDF(n_1^AG) \oplus c_1^A) \oplus c_2^A$

$= H_2(H_2(n_2^A) \oplus c_1^A) \oplus c_2^A$

$= H_2(c_1^A \oplus k^A \oplus c_1^A) \oplus c_2^A$

$= H_2(k^A) \oplus c_2^A$

$= \oplus c_2^A \oplus m^A \oplus c_2^A$

$= m^A$.

### E. Traceability Phase

In the traceability phase, TA can trace IN, which is with misbehaviors, such as leaking data. To be specific, TA can recover $(PID_{IN}, sk_{IN})$ by computing $PID_{IN}, sk_{IN} = c^A \oplus c_2^A$. And then TA is able to find which IN is malicious.

### F. Warrant Update Phase

During the data access of one accident, if IN needs to access more data to better complete digital forensics, then the warrant of it needs to be updated. Due to the warrant is issued by using DIWIm, the warrant update is low overhead. The detailed four steps are as follows:

1) IN sends $\{req_{update}, PID_{IN}\}$ to $\{WI_i\}_{i\in\{1,...,N_U\}}$, which the partial warrant needs to update.

When receiving the request, each $WI_i$ generates the new partial warrant for IN by the following steps:

2) Each $WI_i$ computes $y_i = s_1 + H_3(i)^{1-u_{WI_i}}$, where $u_{WI_i}$ is the updated data access permission identifier of IN stored in $WI_i$.

3) Each $WI_i$ computes $w_{3,i} = 1/s_{2,i}y_i$ and obtains $w_U = (w_{3,i})$.

4) Each $WI_i$ sends $\{enc(w_U, u_{WI_i})_{pk_{IN}}, sig(T)_{sk_{WI_i}}\}$ to IN.

After receiving $\{enc(w_U, u_{WI_i})_{pk_{IN}}, sig(T)_{sk_{WI_i}}\}$, IN conducts the following two steps to obtain the updated warrant:

1) IN computes $w_3 = \prod_{i=1}^{N_U}w_{3,i}$ and $w_{IN,2} = w_2 + w_3 + w_4$.

2) IN obtains the updated warrant $w_{IN} = (w_{IN,1}, w_{IN,2})$ and $\mathbb{A} = \{a_{WI_i}\}_{i=1}^{N_U}$.

So far, IN completes the warrant update, which only needs to update the partial warrant, indicating low overhead.

## V. SECURITY ANALYSIS

In this section, we provide the formal and informal analysis to prove the security attributes introduced in Section III.B.3 and security goals introduced in Section III.B.2.1.

### A. Formal Analysis

**Theorem 1**. If, for any PPT $\mathcal{A}_I$, the probability that $\mathcal{A}_I$ can solve the ECDHP and ECDLP is negligible in any polynomial time, then $Adv_{\mathcal{A}_I}^{LPDP} = |\Pr(win_{\mathcal{A}_I}^{GU})| < \varepsilon$ holds and we claim LPDP satisfies the unforgeability defined by **Definition 1**.

**Theorem 2**. If, for any PPT $\mathcal{A}_{II}$, the probability that $\mathcal{A}_I$ can solve the ECDHP and q-GDHP is negligible in any polynomial time, then $Adv_{\mathcal{A}_{II}}^{LPDP} = |\Pr(win_{\mathcal{A}_{II}}^{GC}) - \dfrac{1}{2}| < \varepsilon$ holds, and we claim LPDP satisfies the confidentiality defined by **Definition 2**.

Due to space limitation, the proofs of **Theorem 1** and **Theorem 2** are given to the supplemental file.

### B. Informal Analysis

This section makes informal analysis of how LPDP achieves the security goals mentioned in Section III.B.2.1. The S1)-S8) listed here corresponds to the S1)-S8) listed in Section III.B.2.1.

**S1)** In the data upload subphase, only someone having $t$ can it generate a valid $sig^A$ used to authenticate itself by RSU. The $t$ is obtained by the DP registration subphase. So, only legitimate DP can be authenticated by RSU. In addition, $sig^A$ can only be verified if $m^A$ has not been unmodified. So, data integrity can be ensured. Besides, $\{\{L_{i,j}^A\}_{j=1}^{N_{WI}-|\mathbb{P}|}, N_{1,i}^A, N_{2,i}^A, c_{1,i}^A, c_{2,i}^A, sig_i^A, R_i^A\}$ sent by DP has the accident identification. So, it can't be replayed.

In the data access subphase, IN can obtain the data only if the data sender knows $PID_{IN}, sk_{IN}$. So, the data origin can be authenticated. Besides, IN can verify the data integrity by verifying if $n_1^{A'}G = n_1^AG$ holds. So, data integrity can be ensured. The timestamp in $\{Enc(req_{data}, PID_{IN}, T)_{pk_{TA}}\}$ can also resist replay attack.

Therefore, LPDP satisfies the data authenticity.

**S2)** In the data access subphase, only IN's $\mathbb{A}$ satisfied $\mathbb{P} \subseteq \mathbb{A}$ can it obtain the valid $n_1^A G$. Then it can use $n_1^A G$ to obtain $m^A$. Otherwise, IN can't decrypt the ciphertext. Therefore, LPDP can protect the data privacy.

**S3)** In the data upload subphase, DP sends $\{\{L_j^A\}_{j=1}^{N_{WI}-|\mathbb{P}|}, N_1^A, N_2^A, c_1^A, c_2^A, sig^A\}$ to RSU to upload data. Since $\{\{L_j^A\}_{j=1}^{N_{WI}-|\mathbb{P}|}, N_1^A, N_2^A, c_1^A, c_2^A, sig^A\}$ doesn't contain any other information of DP itself, such as public key and real identity, attackers can't infer the DP's location from the location of the accident. Therefore, LPDP can protect the DP location privacy.

**S4)** In the warrant issue phase, DP obtains $\{w_{IN_i}, a_{WI_i}\}_{i=1}^{N_{WI}}$ from $\{WI_i\}_{i=1}^{N_{WI}}$. Then it can aggregate $\{w_{IN_i}\}_{i=1}^{N_{WI}}$ into $w_{IN} = (w_{IN,1}, w_{IN,2})$. **Remark 2** presents that $w_{IN}$ can be used to obtain correct $m^A$. Therefore, LPDP satisfies the requirement of the decentralized warrant issuance.

**S5)** In the data access subphase, when IN sends the request, it uses $PID_{IN}$ instead of $ID_{IN}$. Therefore, LPDP satisfies the anonymity requirement.

**S6)** In the data upload subphase, DP sends $\{\{L_j^A\}_{j=1}^{N_{WI}-|\mathbb{P}|}, N_1^A, N_2^A, c_1^A, c_2^A, sig^A, R_i^A\}$ to RSU to upload data. Since $k^A$ and $r_{token}^A$ are randomly selected, attackers can't distinguish between different $\{\{L_j^A\}_{j=1}^{N_{WI}-|\mathbb{P}|}, N_1^A, N_2^A, c_1^A, c_2^A, sig^A\}$. In the data access subphase, IN sends $\{Enc(req_{data}, PID_{IN}, T)_{pk_{TA}}\}$ to TA. Since $PID_{IN}$ is encrypted by ECC, attackers also can't distinguish between different $\{Enc(req_{data}, PID_{IN}, T)_{pk_{TA}}\}$. Therefore, LPDP satisfies the requirement of unlinkability.

**S7)** In the data access subphase, IN sends $\{Enc(req_{data}, PID_{IN}, T)_{pk_{TA}}\}$ to TA. And then TA computes $c^A = c_2^A \oplus (PID_{IN} sk_{IN})$ and returns $\{\mathbb{P}, \{L_i^A\}_{i=1}^{N_{WI}-|\mathbb{P}|}, N_1^A, N_2^A, c_1^A, c^A\}$. If $c^A$ is decrypted and leaked, the leaker must know $PID_{IN}$ and $sk_{IN}$. As a result, TA can trace the malicious IN by using $PID_{IN}$ and $sk_{IN}$. Therefore, LPDP has the traceability feature.

## VI. PERFORMANCE ANALYSIS

In this section, we present the results for performance analysis of LPDP in terms of computation and communication overheads and compare LPDP with [24]-[26] and [28]. Besides, we also conduct the informal analysis for the performance goals introduced in Section III.B.2.2.

To be begin, the experiment setup is introduced. We use JAVA [30] along with the cryptographic tool library JPBC [31]. We choose the elliptic curve to ensure an 80-bit security level. Consequently, the sizes of various parameters are as follows: $|G|$ 256 bits, $|p|$ 160 bits, respectively. We also set $N_{WI} = 5$, $|\mathbb{A}|=|\mathbb{P}|=5$ and $N_U = 1$. The SHA256 function is employed for hashing. The implementation runs on a virtual machine with an Intel(R)_Core(TM)_i7-10700_CPU @2.90GHz and 2GB of RAM, with the Ubuntu 20.04.3 operating system.

### A. Computation Analysis

In this section, we assess the computational overhead of LPDP based on the introduced parameter configuration.

### B. Communication Analysis

This section details the communication overhead of LPDP, based on the previously introduced parameter set. In the warrant issue phase, WIs need to communicate with each other, resulting in additional overhead $(2N_{WI} - 2)|Z_p|$ bits. In the data provision phase, the communication overhead mainly comes from the amount of data to be uploaded or accessed.

### C. Performance Goal Analysis

This section makes analysis of how LPDP achieves the performance goals mentioned in Section III.B.2.2. The P1) and P2) listed here corresponds to the P1) and P2) listed in Section III.B.2.2.

## VII. CONCLUSION

In this paper, we introduce the *L*ightweight and *P*rivacy-preserving *D*ata *P*rovision (LPDP) approach in digital forensics for driverless taxis (DTs). LPDP is composed of three mechanisms to achieve: 1) data authenticity and DP privacy-preserving when uploading data, 2) data and IN privacy-preserving with traceability when accessing data, 3) scalable verification of batch data integrity, and 4) IN warrant update with low overhead. Security analysis and performance evaluation confirm the efficiency of LPDP.

While LPDP performs well in terms of security and performance, it only focuses on the data upload and access phases of vehicle digital forensics (VDF) for DT. Therefore, for the future work, we will further study the problems in other phases.

**Yanwei Gong** is currently pursuing the Ph.D. Degree in Cyberspace Security at Beijing Key Laboratory of Security and Privacy in Intelligent Transportation, Beijing Jiaotong University, China. His interests include identity authentication protocol related to MEC and fully homomorphic encryption acceleration.

**Xiaolin Chang** is currently a professor at the School of Computer and Information Technology, Beijing Jiaotong University, China. Her current research interests include Edge/Cloud computing, Network security, security and privacy in machine learning. She is a senior member of IEEE.

**Jelena Mišić** is Professor of Computer Science at Toronto Metropolitan University, Ontario, Canada. She serves on editorial boards of IEEE Trans. Veh. Technol., Comput. Netw, Ad hoc Netw, Secur. Commun. Netw, Ad Hoc Sens Wirel Netw, Int. J Sens Netw, and Int. J of Telemed Appl. She is a Fellow of IEEE and Member of ACM.

**Vojislav B. Mišić** is Professor of Computer Science at Toronto Metropolitan University, Ontario, Canada. His research interests include performance evaluation of wireless networks and systems and software engineering. He serves on the editorial boards of IEEE Trans. Cloud Comput., Ad hoc Netw, Peer Peer Netw Appl, and Int. J Parallel, Emergent Distrib. Syst.. He is a Senior Member of IEEE and Member of ACM.

**Junchao Fan** is currently pursuing the Ph.D. Degree in Cyberspace Security at Beijing Key Laboratory of Security and Privacy in Intelligent Transportation, Beijing Jiaotong University, China. His interests include trustworthy and secure machine learning and reinforcement learning.

**Kaiwen Wang** is currently pursuing the Ph.D. Degree in Cyberspace Security at Beijing Key Laboratory of Security and Privacy in Intelligent Transportation, Beijing Jiaotong University, China. His interests include network security and secure multi-party computation.